\documentstyle{mn}
\input{psfig.sty}
\newcommand{\etal}{{et al.~}}
\newcommand{\lta}{\la}
\newcommand{\gta}{\ga}

\newcommand{\kmsmpc}{\>{\rm km}\,{\rm s}^{-1}\,{\rm Mpc}^{-1}}

\newcommand{\Mpc}{\>{\rm Mpc}}

\newcommand{\Msun}{\>{\rm M_{\odot}}}
\newcommand{\msun}{\>{\rm M_{\odot}}}
\newcommand{\Lsun}{\>{\rm L_{\odot}}}
\newcommand{\MLsun}{\>({\rm M}/{\rm L})_{\odot}}

\newcommand{\beq}{\begin{equation}}
\newcommand{\eeq}{\end{equation}}

\newcommand{\mpch}{\>h^{-1}{\rm {Mpc}}}

\newcommand{\msunh}{\>h^{-1}\rm M_\odot}

\newcommand{\walpha}{\tilde{\alpha}}
\newcommand{\wLstar}{\tilde{L}^{*}}

\newcommand{\apj}{ApJ}

\newcommand{\aj}{AJ}
\newcommand{\mnras}{MNRAS}

\newdimen\hssize
\newdimen\hdsize 
\begin{document}

%%%%%%%%%%%%%%%%%%%%%%%%%%%%%%%%%%%%%%%%%%%%%%%%%%%%%%%%%%%%%%%%%%%%%%%%%%

\title[The Dependence of the Galaxy Luminosity Function on Large-Scale Environment]
      {The Dependence of the Galaxy Luminosity Function on Large-Scale Environment}
\author[Mo, Yang, van den Bosch \& Jing]
       {H.J. Mo$^{1}$, Xiaohu Yang$^{1,2}$, Frank C. van den
      Bosch$^{3,4}$, Y.P Jing$^5$ 
      \thanks{E-mail: hjmo@nova.astro.umass.edu}\\
      $^1$ Department of Astronomy, University of Massachusetts,
           Amherst MA 01003-9305, USA\\
      $^2$ Center for Astrophysics, University of Science and Technology
           of China, Hefei, Anhui 230026, China\\
      $^3$ Institute of Astronomy, Swiss Federal Institute of
           Technology, ETH H\"onggerberg, CH-8093, Zurich,
           Switzerland\\ 
      $^4$ Max-Planck-Institut f\"ur Astrophysik
          Karl-Schwarzschild-Strasse 1, 85748 Garching, Germany\\
      $^5$ Shanghai Astronomical Observatory; the Partner Group of MPA,
       Nandan Road 80,  Shanghai 200030, China}
%%%%%%%%%%%%%%%%%%%%%%%%%%%%%%%%%%%%%%%%%%%%%%%%%%%%%%%%%%%%%%%%%%%%%%%%%%

\date{}

%\pubyear{2003}

\maketitle

\label{firstpage}

%%%%%%%%%%%%%%%%%%%%%%%%%%%%%%%%%%%%%%%%%%%%%%%%%%%%%%%%%%%%%%%%%%%%%%%%%%

\begin{abstract}
  A  basic assumption  in current  halo occupation  model is  that the
  properties of  a galaxy depend only  on the mass of  its dark matter
  halo.  An important  consequence of this is that  the segregation of
  the galaxy population by  large-scale environment is entirely due to
  the environmental dependence of  the halo population.  In this paper
  we use  such a model to  predict how the  galaxy luminosity function
  depends on  large-scale environment.   The latter is  represented by
  the density contrast ($\delta$)  averaged over a spherical volume of
  radius $R=8\mpch$.   The model predicts that  the Schechter function
  is  a good  approximation to  the luminosity  functions  of galaxies
  brighter than $\sim 10^9  h^{-2}\Lsun$ ($b_j$-band) in virtually all
  environments.  The characteristic luminosity, $L^{\star}$, increases
  moderately  with $\delta$.   The faint-end  slope, $\alpha$,  on the
  other  hand,  is  quite  independent  of  $\delta$.   However,  when
  splitting the  galaxy population  into early and  late types,  it is
  found that  for late-types  $\alpha$ is virtually  constant, whereas
  for  early-types  $\alpha$ increases  from  $\sim  -0.3$ in  underdense
  regions  ($\delta  \sim -0.5$)  to  $\sim  -0.8$  in highly  overdense
  regions with $\delta \sim  10$. The luminosity function at $L_{b_j}<
  10^9 h^{-2}\Lsun$ is significantly steeper than the extrapolation of
  the  Schechter  function  that  fits the  brighter  galaxies.   This
  steepening is  more significant  for early-types and  in low-density
  environments.  The  model also predicts that  the luminosity density
  and mass  density are closely correlated.  The  relation between the
  two is monotonic but highly  non-linear.  This suggests that one can
  use the  luminosity density, averaged  over a large volume,  to rank
  the mass  density. This, in  turn, allows the  environmental effects
  predicted here to be tested by observations.
\end{abstract}

%%%%%%%%%%%%%%%%%%%%%%%%%%%%%%%%%%%%%%%%%%%%%%%%%%%%%%%%%%%%%%%%%%%%%%%%%%


%%%%%%%%%%%%%%%%%%%%%%%%%%%%%%%%%%%%%%%%%%%%%%%%%%%%%%%%%%%%%%%%%%%%%%%%%%

\section{Introduction}

In a  hierarchical cosmogony  like the cold  dark matter  (CDM) model,
galaxies are assumed to form in dark matter haloes (e.g. White \& Rees
1978).  A generic prediction of  such cosmogony is that the properties
of the galaxy  population must be closely related to  that of the halo
population.  If the cosmological density  field is Gaussian and if the
power of density perturbations extends to large scales, as is the case
in  the current  CDM cosmogony,  halo  properties are  expected to  be
correlated to some  degree with the large-scale structure,  and so the
properties of the  galaxy population are also expected  to change with
large-scale  environment.  With high-resolution  numerical simulations
much has been learned about the  halo population.  It turns out that a
strong correlation  with the  large-scale environment exists  only for
halo  mass; all  other important  halo properties  are at  most weakly
correlated with  the large-scale environment  at any given time  (e.g. 
Lemson \&  Kauffmann, 1999).  This  result, combined with  our current
model of  galaxy formation, implies that  the environmental dependence
of the galaxy population is mainly  due to the change of the halo mass
function  with  large-scale environment.   We  can  therefore hope  to
understand the  environmental dependence  of the galaxy  population by
understanding  how galaxies occupy  dark haloes  of different  masses. 
This is  indeed the  approach taken by  the so-called  halo occupation
models, in which  the {\it number} of galaxies per  halo is assumed to
depend  on halo  {\it mass}  only (e.g.   Jing, Mo  \&  B\"orner 1998;
Peacock \& Smith 2000; Seljak  2000; Scoccimarro et al. 2001; 
Benson 2001; Bullock,
Wechsler \&  Sommerville 2002; Berlind \&  Weinberg 2002; Kang  et al. 
2002; Zheng  et al.  2002; Yang, Mo  \& van  den Bosch 2003a;  van den
Bosch, Yang \& Mo 2003; Magliocchetti \& Porciani 2003; Scranton 2002;
Zehavi \etal 2003; Yan, Madgwick \& White 2003; Kravtsov et al. 2003).
Note that  this assumption is  non-trivial, as it implies  that galaxy
formation is largely a local process in individual dark matter haloes.
This assumption was also made in many of the semi-analytic models
based on Monte-Carlo halo merging trees 
(e.g. Kauffmann, White \& Guiderdoni 1993;
Cole et al. 1994; Sommerville et al. 1999), because in such models
the merging histories are statistically the same for all haloes 
of the same mass at a given time 
Berlind et al. (2003) found that this assumption is consistent with 
their numerical simulations. 
 
To test  the validity of this  assumption, we use  the halo occupation
model recently developed by Yang  \etal (2003a), which is based on the
conditional luminosity  function (hereafter  CLF) of galaxies  in dark
haloes of  given mass, to  predict how the galaxy  luminosity function
(hereafter LF)  changes with large-scale environment.   In this model,
the change of  the LF with large-scale environment  is entirely due to
the change  of the  halo mass function  with large-scale  environment. 
Since  the conditional  mass function  of  dark matter  haloes can  be
accurately  obtained from  cosmological $N$-body  simulation,  the CLF
model can be  used to make accurate predictions for  how the galaxy LF
changes  with large-scale  density  field.  We  also  show that,  when
averaged over a large  volume, the predicted galaxy luminosity density
is closely correlated with the underlying mass density. Therefore, the
luminosity density can be used to rank the mass density, and the model
predictions presented  here can be tested using  large galaxy redshift
surveys, such as the  two-degree Field Galaxy Redshift Survey (2dFGRS;
Colless \etal  2001) and the Sloan  Digital Sky Survey  (SDSS; York et
al.  2000).

The paper  is arranged as follows. In  Section~\ref{sec:clf} we review
the CLF model and  describe how it can be used to  calculate the LF of
galaxies  in different environments.   Section~\ref{sec:main} presents
the model predictions  for how the galaxy LF  changes with large-scale
density field.   In Section~\ref{sec:tests} we use our  model to study
the  correlation  between  the   galaxy  luminosity  density  and  the
underlying mass  density, and discuss  how the predictions of  our CLF
formalism can  be tested with observations.  We  summarize our results
in Section~\ref{sec:discussion}.

\section{From the conditional luminosity function
to galaxy luminosity function}
\label{sec:clf}

In Yang, Mo \& van  den Bosch (2003a, hereafter Paper~I), we developed
a  formalism, based  on  the conditional  luminosity function  $\Phi(L
\vert M)$, to link the distribution of galaxies to that of dark matter
haloes.   We introduced  a parameterized  form for  $\Phi(L  \vert M)$
which  we constrained  using the  LF  and the  correlation lengths  as
function  of  luminosity.   In  van  den Bosch,  Yang  \&  Mo  (2003a,
hereafter Paper  II), we extended  our model by  constructing separate
CLFs  for  the  early-  and  late-type galaxies.   The  CLF  formalism
developed  in   these  two  papers  has  subsequently   been  used  to
investigate galaxy clustering as function of luminosity and type (Yang
\etal 2003b),  to put constraints on cosmological  parameters (van den
Bosch, Mo \& Yang 2003b),  to constrain redshift-evolution in the halo
model  (Yan,  Madgwick  \&   White  2003),  and  to  characterize  the
population of satellite galaxies (van  den Bosch \etal 2004).  In this
paper, we use the CLF formalism to predict how the luminosity function
of galaxies depends on  large-scale environment.  For completeness, we
briefly summarize  in the  following the main  ingredients of  the CLF
formalism,  and  we refer  the  reader  to  papers~I and~II  for  more
details.

The conditional  luminosity function $\Phi(L \vert M)  {\rm d}L$ gives
the average number  of galaxies with luminosities in  the range $L \pm
{\rm d}L/2$ that reside in haloes  of mass $M$. It is parameterized by
a Schechter function:
\begin{equation}
\label{phiLM}
\Phi(L  \vert  M)  {\rm  d}L  = {\tilde{\Phi}^{*}  \over  \wLstar}  \,
\left({L \over  \wLstar}\right)^{\walpha} \, \,  {\rm exp}(-L/\wLstar)
\, {\rm d}L,
\end{equation}
where   $\wLstar   =   \wLstar(M)$,   $\walpha   =   \walpha(M)$   and
$\tilde{\Phi}^{*}  = \tilde{\Phi}^{*}(M)$  are all  functions  of halo
mass  $M$. Following Papers~I  and~II, we write the  average total 
mass-to-light ratio of a halo with mass $M$ as
\begin{equation}
\label{MtoLmodel}
\left\langle {M \over L} \right\rangle(M) = {1 \over 2} \,
\left({M \over L}\right)_0 \left[ \left({M \over M_1}\right)^{-\gamma_1} +
\left({M \over M_1}\right)^{\gamma_2}\right],
\end{equation}
which has four free parameters: a characteristic mass $M_1$, for which
the  mass-to-light  ratio  is  equal  to $(M/L)_0$,  and  two  slopes,
$\gamma_1$ and  $\gamma_2$, that specify the behavior  of $\langle M/L
\rangle$  at the  low and  high  mass ends,  respectively.  A  similar
parameterization   is  adopted   for  the   characteristic  luminosity
$\wLstar(M)$:
\begin{equation}
\label{LstarM}
{M \over \wLstar(M)} = {1 \over 2} \, \left({M \over L}\right)_0 \,
f(\walpha) \, \left[ \left({M \over M_1}\right)^{-\gamma_1} +
\left({M \over M_2}\right)^{\gamma_3}\right],
\end{equation}
with
\begin{equation}
\label{falpha}
f(\walpha) = {\Gamma(\walpha+2) \over \Gamma(\walpha+1,1)}.
\end{equation}
Here  $\Gamma(x)$   is  the  Gamma  function   and  $\Gamma(a,x)$  the
incomplete Gamma  function.  This parameterization  has two additional
free  parameters: a characteristic  mass $M_2$  and a  power-law slope
$\gamma_3$.   For $\walpha(M)$ we  adopt a  simple linear  function of
$\log(M)$,
\begin{equation}
\label{alphaM}
\walpha(M) = \alpha_{15} + \eta \, \log(M_{15}),
\end{equation}
with $M_{15}$ the halo mass in units of $10^{15} \msunh$, $\alpha_{15}
= \walpha(M_{15}=1)$, and $\eta$ describes the change of the faint-end
slope  $\walpha$ with  halo  mass.  Note  that  once $\walpha(M)$  and
$\wLstar  (M)$ are  given, the  normalization of  the  conditional LF,
$\tilde{\Phi}^{*}(M)$,  is  obtained  through  equations~(\ref{phiLM})
and~(\ref{MtoLmodel}),  using  the   fact  that  the  total  (average)
luminosity in a halo of mass $M$ is
\begin{equation}
\label{meanL}
\langle L \rangle(M) = \int_{0}^{\infty}  \Phi(L \vert M) \, L
\, {\rm d}L = \tilde{\Phi}^{*} \, \wLstar \, \Gamma(\walpha+2).
\end{equation}
Finally, we introduce the mass  scale $M_{\rm min}$ below which we set
the CLF to zero; i.e., we assume that no stars form inside haloes with
$M  < M_{\rm  min}$.   Motivated by  reionization considerations  (see
Paper~I  for details)  we adopt  $M_{\rm min}  = 10^{9}  h^{-1} \Msun$
throughout.

In order  to split the galaxy  population in early and  late types, we
follow Paper~II and introduce  the function $f_{\rm late}(L,M)$, which
specifies the  fraction of galaxies  with luminosity $L$ in  haloes of
mass  $M$  that are  late-type.   The  CLFs  of late-  and  early-type
galaxies are then given by
\begin{equation}
\label{CLFl}
\Phi_{\rm late}(L \vert M) {\rm d}L = f_{\rm late}(L,M) \, \Phi(L
\vert M) {\rm d}L
\end{equation}
and
\begin{equation}
\label{CLFe}
\Phi_{\rm early}(L \vert M) \, {\rm d}L = \left[ 1 - f_{\rm late}(L,M)
\right] \, \Phi(L \vert M) \, {\rm d}L\,.
\end{equation}
As  with the  CLF for  the entire  population of  galaxies, $\Phi_{\rm
late}(L \vert M)$ and $\Phi_{\rm early}(L \vert M)$ are constrained by
2dFGRS measurements of the LFs and the correlation lengths as function
of   luminosity.    We  assume   that   $f_{\rm   late}(L,M)$  has   a
quasi-separable form
\begin{equation}
\label{fracdef}
f_{\rm late}(L,M) = g(L) \, h(M) \, q(L,M).
\end{equation}
Here
\begin{equation}
\label{qlm}
q(L,M) = \left\{
\begin{array}{lll}
1                      & \mbox{if $g(L) \, h(M) \leq 1$} \\
{1 \over g(L) \, h(M)} & \mbox{if $g(L) \, h(M) > 1$}
\end{array} \right.
\end{equation}
is to ensure that $f_{\rm late}(L,M) \leq 1$. We adopt
\begin{equation}
\label{gl}
g(L) = {\hat{\Phi}_{\rm late}(L) \over \hat{\Phi}(L)}
{\int_{0}^{\infty} \Phi(L \vert M) \, n(M) \, {\rm d}M \over
 \int_{0}^{\infty} \Phi(L \vert M) \, h(M) \, n(M) \, {\rm d}M}
\end{equation}
where $n(M)$ is  the halo mass function (Sheth  \& Tormen 1999; Sheth,
Mo  \& Tormen  2001), $\hat{\Phi}_{\rm  late}(L)$  and $\hat{\Phi}(L)$
correspond  to the  {\it observed}  LFs  of the  late-type and  entire
galaxy samples, respectively, and
\begin{equation}
\label{hm}
h(M) = \max \left( 0, \min\left[ 1, \left({{\rm log}(M/M_a)
\over {\rm log}(M_b/M_a)} \right) \right] \right)
\end{equation}
with $M_a$  and $M_b$ two  additional free parameters, defined  as the
masses at which $h(M)$ takes  on the values $0$ and $1$, respectively.
As shown  in Paper~II, this parameterization allows  the population of
galaxies  to  be  split  in  early- and  late-types  such  that  their
respective LFs and clustering properties are well fitted.

In  Papers~I  and~II we  presented  a  number  of different  CLFs  for
different  cosmologies and  different assumptions  regarding  the free
parameters.   In  what  follows  we  focus on  the  flat  $\Lambda$CDM
cosmology with  $\Omega_m=0.3$, $\Omega_{\Lambda}=0.7$ and $h=H_0/(100
\kmsmpc) = 0.7$  and with initial density fluctuations  described by a
scale-invariant  power  spectrum  with normalization  $\sigma_8=0.9$.  
These cosmological parameters are in  good agreement with a wide range
of  observations, including  the  recent WMAP  results (Spergel  \etal
2003),  and in  what follows  we refer  to it  as  the ``concordance''
cosmology.  Finally,  we adopt the CLF with  the following parameters:
$M_1  =  10^{10.94}  h^{-1}  \Msun$,  $M_2=10^{12.04}  h^{-1}  \Msun$,
$M_a=10^{17.26}   h^{-1}   \Msun$,   $M_b=10^{10.86}   h^{-1}   \Msun$,
$(M/L)_0=124    h     \MLsun$,    $\gamma_1=2.02$,    $\gamma_2=0.30$,
$\gamma_3=0.72$,  $\eta=-0.22$ and  $\alpha_{15}=-1.10$.  As  shown in
paper~II, this model (referred to as model~D) yields excellent fits to
the observed LFs  and the observed correlation lengths  as function of
both luminosity and type.  We emphasize, however, that our results are
not sensitive to  uncertainties in the CLF; had we  chosen models A, B
or C in paper~II, instead of model D, the results presented below 
would have been virtually identical.
\begin{figure}
\centerline
{\psfig{figure=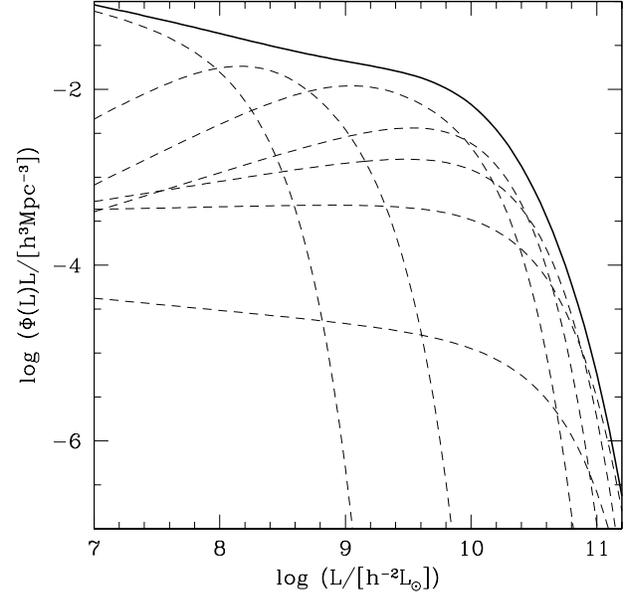,width=0.5\hdsize}}
\caption{The  contribution  to the  total  luminosity function  (solid
  curve)  by  haloes  in  various mass  ranges:  $M/[h^{-1}\msun]  \le
  5\times  10^{10}$,  $5\times   10^{10}$  -  $10^{11}$,  $10^{11}$  -
  $10^{12}$, $10^{12}$ - $10^{13}$, $10^{13}$ - $10^{14}$, $10^{14}$ -
  $10^{15}$, $>10^{15}$ (broader curve corresponds to larger mass).}
\label{fig:LFasM}
\end{figure}
Since $\Phi  (L\vert M)$  is the average  number of galaxies  per unit
luminosity in  a halo  with mass  $M$, one can  obtain the  galaxy LF,
$\Phi(L)$, from the halo mass function, $n(M)$, according to
\begin{equation}
\Phi(L)= \int \Phi(L \vert M) \, n(M) \, {\rm d}M. 
\end{equation}
Fig.\ref{fig:LFasM} shows how haloes of different masses contribute to
the total luminosity function.  Note that the shape of the CLF changes
significantly with  halo mass,  and that  the faint end  of the  LF is
dominated by galaxies hosted by low-mass haloes.
 
If we make the assumption that the CLF is statistically independent of
the large scale environment, the galaxy LF in a region of
overdensity $\delta$ follows from
\begin{equation}
\Phi(L \vert \delta) = \int \Phi(L\vert M) \, n(M \vert \delta) \,
{\rm d}M.
\end{equation}
Here $n(M\vert\delta)$ is the conditional mass function of dark matter
haloes, which gives the number density of haloes as a function of halo
mass in  an environment with  average mass overdensity  $\delta \equiv
[{\cal M}-{\overline  {\cal M}}]  /{\overline {\cal M}}$  (with ${\cal
  M}$ the  total mass  in volume $V$,  and ${\overline {\cal  M}}$ the
mean  mass in  all  volumes $V$).   Thus,  in the  CLF formalism,  the
$\delta$-dependence  of   the  galaxy   LF  enters  only   though  the
conditional mass function $n(M\vert \delta)$.
\begin{figure*}
\centerline{\psfig{figure=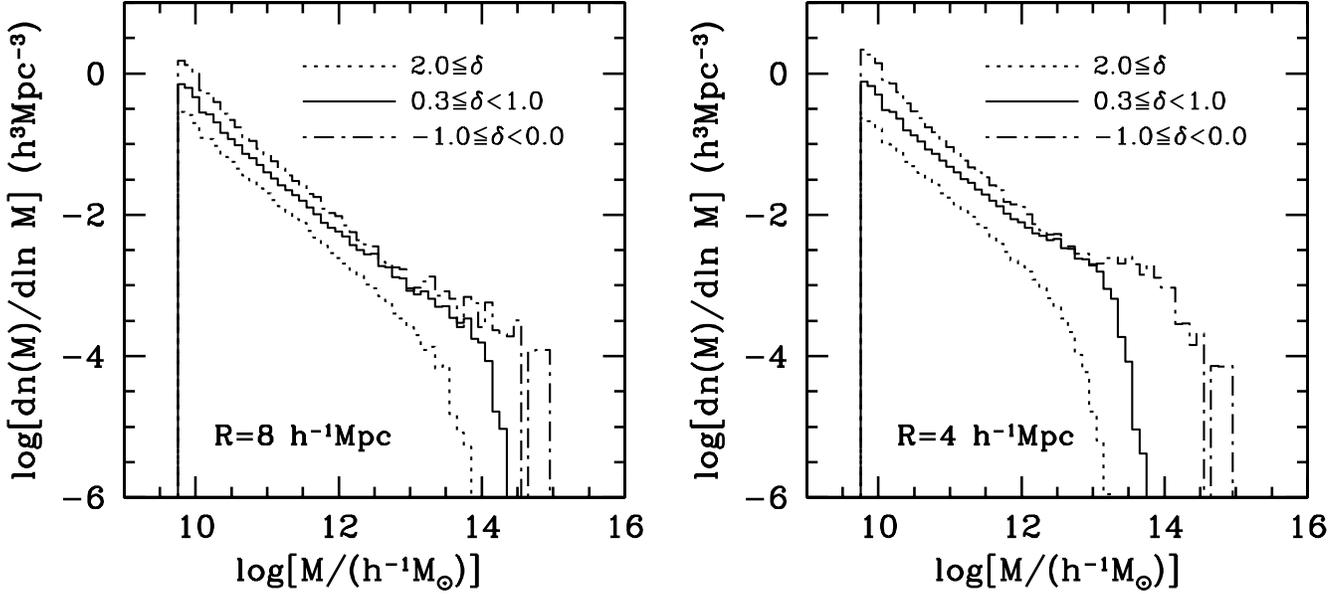,width=1.0\hdsize}}
\caption{The conditional mass functions of dark matter haloes in various
  environments,  specified  by  the  mass overdensity  $\delta$  in  a
  spherical volume  with radius $R=  8 h^{-1} \Mpc$  (left-hand panel)
  and $R= 4 h^{-1} \Mpc$ (right-hand panel).}
\label{fig:conditionalMF}
\end{figure*}

Fig.\,\ref{fig:conditionalMF} shows the conditional mass functions for
haloes  in several representative  environments.  These  functions are
derived from  a high-resolution $N$-body simulation  which follows the
motions of  $512^3$ particles with a  P$^3$M code in  a $100\mpch$ box
(see  Jing  \&  Suto  2002  for details),  assuming  the  $\Lambda$CDM
`concordance'  cosmology  specified  above.   Dark matter  haloes  are
identified  using  the standard  friends-of-friends  algorithm with  a
linking length of $0.2$  times the mean inter-particle separation.  As
one  can see,  the shape  of the  conditional mass  function  is quite
independent of  $\delta$ at the low  mass end, while the  shape at the
high-mass  end depends significantly  on $\delta$.   The break  in the
conditional mass function at the  high-mass end occurs at a lower mass
for  lower $\delta$.   This reflects  the fact  that the  formation of
massive haloes is suppressed in low  density regions due to bias (e.g. 
Mo \& White 1996; Gottloeber et  al. 2003).  For a given $\delta$, the
break occurs at smaller mass if  the volume used to define $\delta$ is
smaller.

Ideally,  if  we have  an  accurate  model  for the  conditional  mass
function, we  can combine it with the  conditional luminosity function
to construct  an analytical model  for the $\delta$-dependence  of the
galaxy  LF.    Unfortunately,  the  model   based  on  peak-background
splitting (e.g. Cole \& Kaiser 1989; Mo \& White 1996) 
is not sufficiently accurate for our
purpose, even if ellipsoidal collapse is taken into account (Sheth, Mo
\& Tormen 2001;  Sheth \& Tormen 2002).  The reason  for this is that,
when the  total mass contained in  a volume becomes  comparable to the
mass of individual halos in  consideration, as is the case for massive
halos  in low-density volumes,  the peak-background  splitting becomes
inaccurate (e.g.  Sheth \& Tormen  2002).  We therefore decided to use
the  simulated  samples constructed  in  Yang  \etal  (2003b) for  our
purpose.   These simulated  samples were  obtained by  populating dark
matter haloes  in $N$-body simulations with galaxies  according to the
CLF model  described above.  Here  we use the sample  constructed from
one  of  the $100\mpch$  simulations,  which  is  complete down  to  a
$b_j$-band  luminosity of  $L_{b_j} \sim  10^8 h^{-2}\Lsun$  (see Yang
\etal 2003b for details).

\section{Galaxy luminosity function in different environments}
\label{sec:main}

\begin{figure}
\centerline
{\psfig{figure=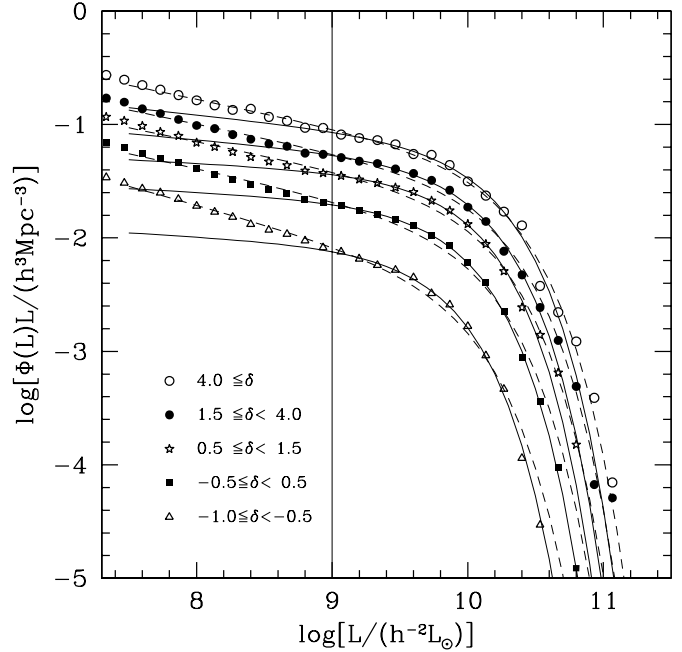,width=0.5\hdsize}}
\caption{Galaxy luminosity functions in regions with different mass 
  overdensities:  $-1.0 \leq  \delta <  -0.5$ (open  triangles), $-0.5
  \leq  \delta  <  0.5$  (solid  squares), $0.5  \leq  \delta  <  1.5$
  (asterisks), $1.5  \leq \delta <  4.0$ (solid circles),  and $\delta
  \geq 4.0$ (open circles).  The solid and dashed curves correspond to
  the  best-fit Schechter  functions, fit  over the  luminosity ranges
  $L_{b_J} \geq  10^9 h^{-2} \Lsun$ and $L_{b_J}  \geq 10^{7.5} h^{-2}
  \Lsun$, respectively.}
\label{fig:conLF_all}
\end{figure}

Fig.~\ref{fig:conLF_all}  shows  the LFs  of  galaxies  in regions  of
different  mass  overdensities  $\delta$  (defined over  spheres  with
radius $R = 8 h^{-1} \Mpc$). As one can see, the {\it shape} of the LF
at the faint end is quite independent of $\delta$.  This is due to the
fact that most faint galaxies reside in low mass haloes, for which the
{\it shape} of  the conditional mass function $n(M  | \delta)$ is only
weakly  dependent on  $\delta$  (cf. Fig.   2).   The most  pronounced
difference between  LFs in different mass density  environments is the
break  at the  bright end,  which  occurs at  fainter luminosities  in
lower-density regions.  This,  in turn, is simply a  reflection of the
fact that most bright galaxies reside in massive haloes.

Our  CLF   model  also  predicts  a  pronounced   difference  for  the
environment dependence  of the LFs  of early- and late-type  galaxies. 
As shown in  Fig. 4, the {\it shape} of the  LF of early-type galaxies
depends  rather strongly  on  $\delta$.  For  the late-type  galaxies,
however,  no  such  pronounced  shape-dependence is  predicted.   Once
again, this behaviour is easy  to understand from the conditional halo
mass function  and the CLF:  many of the fainter  early-types actually
reside in clusters (see Paper~II), whose abundance depends strongly on
$\delta$. The majority  of the faint late-type galaxies,  on the other
hand, reside in relatively low mass haloes, for which the shape of the
conditional halo  mass function is roughly  independent of environment
(cf. Fig.2).
\begin{figure*}
\centerline
{\psfig{figure=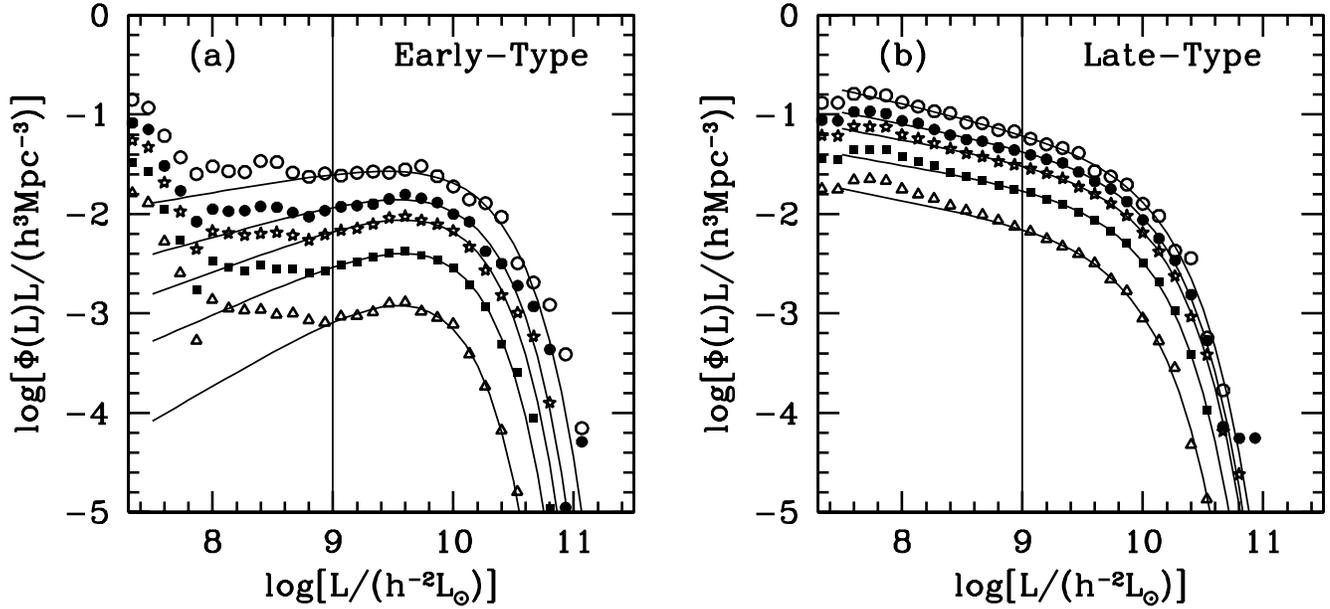,width=\hdsize}}
\caption{The  same  as  Fig.~\ref{fig:conLF_all}, but  for  (a) 
  early-type and  (b) late-type  galaxies. Solid curves  correspond to
  the best-fit  Schechter functions, fit over the  range $L\geq 10^{9}
  h^{-2}\Lsun$.}
\label{fig:conLF_type}
\end{figure*}
\begin{figure*}
\centerline
{\psfig{figure=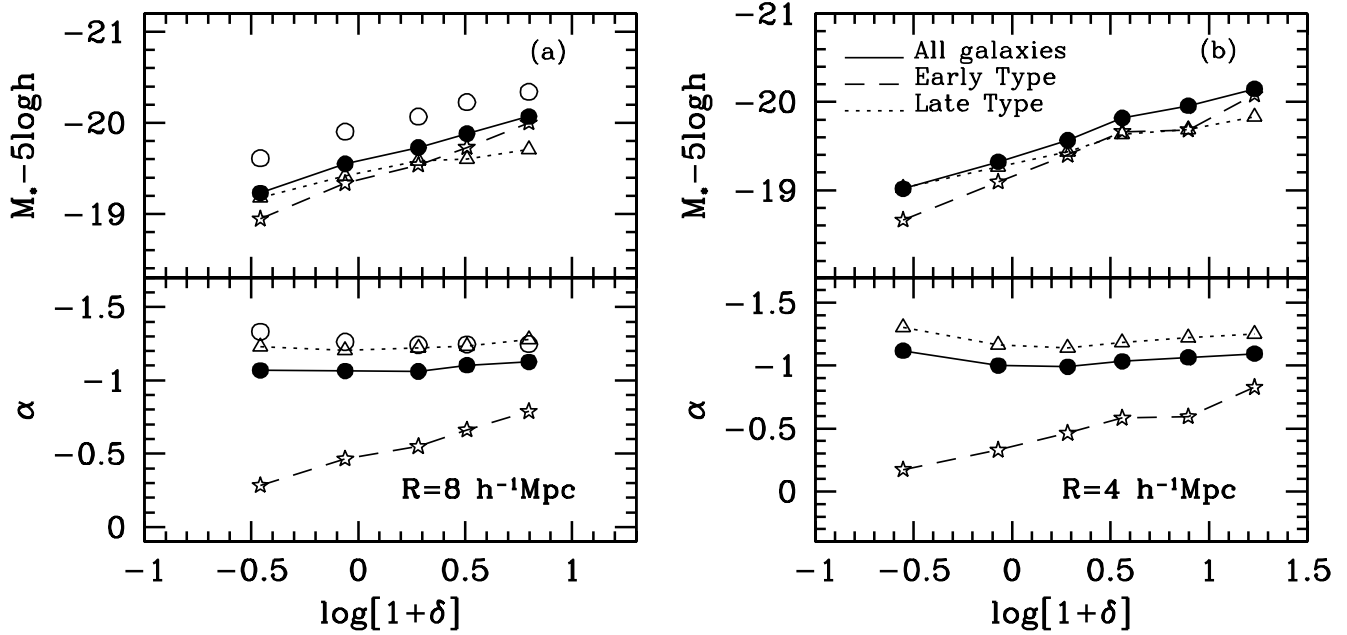,width=\hdsize}}
\caption{The characteristic luminosity and faint-end slope $\alpha$ as
  a  function of  $\delta$,  defined as  the  mean overdensity  within
  individual  spheres of  radius $R=8\mpch$  (left panel)  and $R  = 4
  \mpch$ (right  panel).  Symbols connected with lines  are results of
  the fit  over the luminosity  range $L \geq 10^{9}h^{-2}  \Lsun$ for
  the  total  (solid line),  early-type  (dashed  line) and  late-type
  (dotted  line) samples.   For comparison,  the open  circles  in the
  left-hand panel show  the results for the total  sample when fitting
  over  the more  extended luminosity  range $L  \geq  10^{7.5} h^{-2}
  \Lsun$. }
\label{fig:LF_para}
\end{figure*}

To further  quantify how  the LF  depends on the  mean density  of the
environment, we  fit each of  the luminosity functions by  a Schechter
form.  For  most of  our discussion, we  only fit over  the luminosity
range $L \geq 10^{9}h^{-2}  \Lsun$ ($M_{b_j}-5\log h \lta -17.2$).  In
real redshift surveys, such as  the 2dFGRS, galaxies fainter than this
are observed  only within a  relatively small local volume,  making it
difficult  to study  their large-scale  environment  dependencies.  As
shown   in  Figs.~\ref{fig:conLF_all}   and~\ref{fig:conLF_type},  the
Schechter function provides in general a reasonable fit to the LF over
a large  range in luminosity.   This is not  a trivial result,  as the
shape  of  the  CLF  varies  significantly  with  halo  mass  and  the
conditional mass  function at the high-mass  end changes significantly
with $\delta$.

\begin{figure*}
\centerline
{\psfig{figure=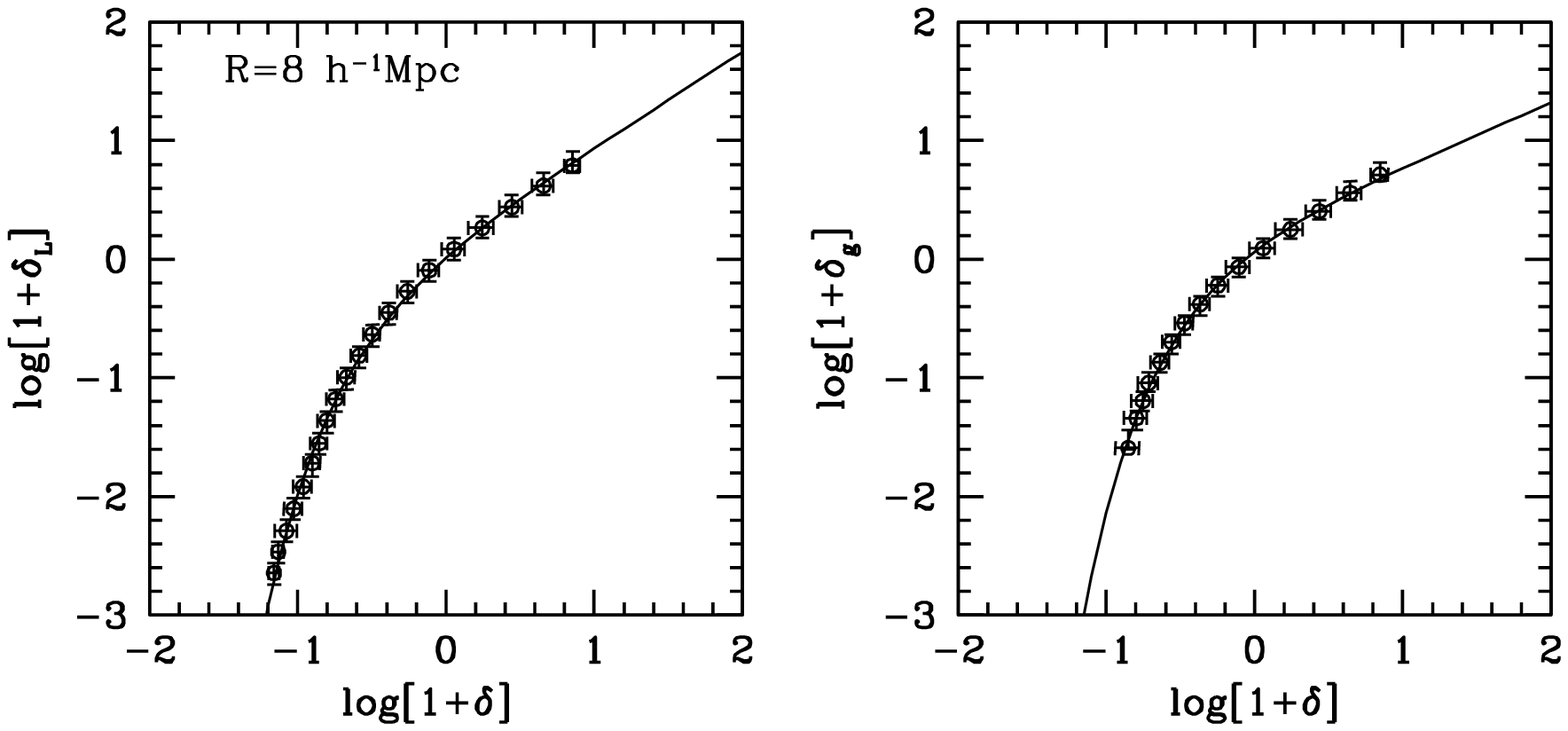,width=\hdsize}}
\caption{The left panel shows the relation  between luminosity density 
  ($\delta_L$,   calculated  using  all   galaxies  resolved   in  the
  simulation) and mass density ($\delta$), while the right panel shows
  the  galaxy number  density ($\delta_g$,  calculated  using galaxies
  with luminosity  $L\ge 10^9  h^{-2}\Lsun$).  All quantities  are the
  averages within a sphere of radius $R=8\mpch$, and normalized to the
  corresponding mean  densities of the universe.  The  curves show the
  fits  to the  relations. Error  bars for  $\delta_L$  and $\delta_g$
  represent the ranges of the quantities, while errorbars for $\delta$
  are $1\sigma$ standard deviations.   Note that below a certain value
  of    $\delta$,   $\log(1+\delta_g)$   becomes    $-\infty$,   while
  $\log(1+\delta_L)$ is still finite. }
\label{fig:dlasdm}
\end{figure*}

The  left  panel of  Fig.~\ref{fig:LF_para}  shows the  characteristic
luminosity  $L^*$ and  the faint  end slope  $\alpha$ as  functions of
$\delta$. In  all cases, the characteristic  luminosity increases with
$\delta$, reflecting  the fact that massive haloes,  which host bright
galaxies,  are preferentially  located in  high-density  regions.  The
increase  is only  moderate:  for the  total  sample it  is about  one
magnitude  from  the lowest  to  the  highest  densities probed.   The
increase  is  more  significant   for  early-type  galaxies  than  for
late-type  galaxies.   For late-type  galaxies,  the increase  becomes
insignificant for $\delta\ga 1$.

For the  total population  the faint-end slope  $\alpha$ is  roughly a
constant,  with a  value about  $-1.1$.  The  faint-end slope  is also
quite independent of $\delta$ for late-type galaxies, with $\alpha\sim
-1.25$.   In contrast,  the  faint-end slope  for early-type  galaxies
changes  from  $\alpha  \sim  -0.3$  for  low-density  regions  to
$\alpha\sim -0.8$ for  $\delta\sim  10$. As  mentioned  above, this  is
mainly due to  the fact that a large number  of relatively faint early
types reside in massive haloes whose mass function depends strongly on
$\delta$.

If we fit  the luminosity function over the entire  range with $L \gta
10^{7.5}h^{-2}  \Lsun$ ($M_{b_j}-5\log h  \lta -13.5$),  the faint-end
slope  becomes steeper for  low-density regions  (cf.  open  and solid
circles in  left-panel of  Fig.~\ref{fig:LF_para}.  Note that  in this
case,  the   Schechter  function  is   no  longer  a  good   fit  (see
Fig.~\ref{fig:conLF_all}) for low-density regions; the faint-end slope
is much  steeper than  that of the  extrapolation from  the luminosity
function at the brighter end.   This departure starts at $L\sim 10^{9}
h^{-2}\Lsun$ and is more significant for early-type galaxies.

We have also made similar analyses for spherical volumes with a radius
of $4\mpch$, the results of which are shown in the right-hand panel of
Fig.~\ref{fig:LF_para}.  A noticeable difference with using volumes of
$8  \mpch$  radius is  that  the  characteristic luminosity  increases
faster with $\delta$. The reason is that the conditional mass function
at the  high-mass end  reveals a stronger  dependence on  $\delta$ for
smaller radius $R$ (see Fig.\,\ref{fig:conditionalMF}).

\section{Predictions for observational tests}
\label{sec:tests}

So far  we have  investigated how  the galaxy LF  depends on  the mass
density averaged  over spherical  volumes of $8  h^{-1} \Mpc$  radius. 
Unfortunately,  it  is not  easy  observationally  to obtain  accurate
measures for  the mass density  field in the Universe.   Therefore, in
order to facilitate an observational test of the predictions presented
above, we  need to  find a quantity  that is  (i) easy to  derive from
observations, and (ii) that can be used to rank the mass density.

When  averaged  within  a  large  volume, the  {\it  number}  density,
$\delta_{\rm g}$ of galaxies  (with luminosities above some value) and
the luminosity density, $\delta_L$, are both expected to be correlated
with  the underlying  mass  density.  Such  relations  can be  derived
directly from  the simulated catalogs  constructed with the  CLF (Yang
\etal 2003b).   Fig.~\ref{fig:dlasdm} shows  the results.  As  one can
see, there  is a tight correlation between  the luminosity overdensity
$\delta_L$  and   mass  overdensity  $\delta$.    Here  $\delta_L$  is
calculated using all galaxies with  $L \ge L_{\rm min} = 10^{7} h^{-2}
\Lsun$, but it is not sensitive to this lower limit as long as $L_{\rm
  min} \lta 10^{9} h^{-2} \Lsun$.   Thus the luminosity density can be
used to rank the mass density.  The use of galaxy {\it number} density
is  trickier,  because the  scatter  is  quite  large for  low-density
volumes (for  example, $\delta_{\rm g}$, here defined  by all galaxies
with $L\ge 10^{9} h^{-2} \Lsun$,  becomes zero at $\delta<-0.8$ in the
right panel  of Fig.\,\ref{fig:dlasdm}) and because it  depends on the
luminosity range used in  defining the galaxy overdensity $\delta_{\rm
  g}$ (since the number density  of galaxies is dominated by the faint
ones).

\begin{figure}
\centerline
{\psfig{figure=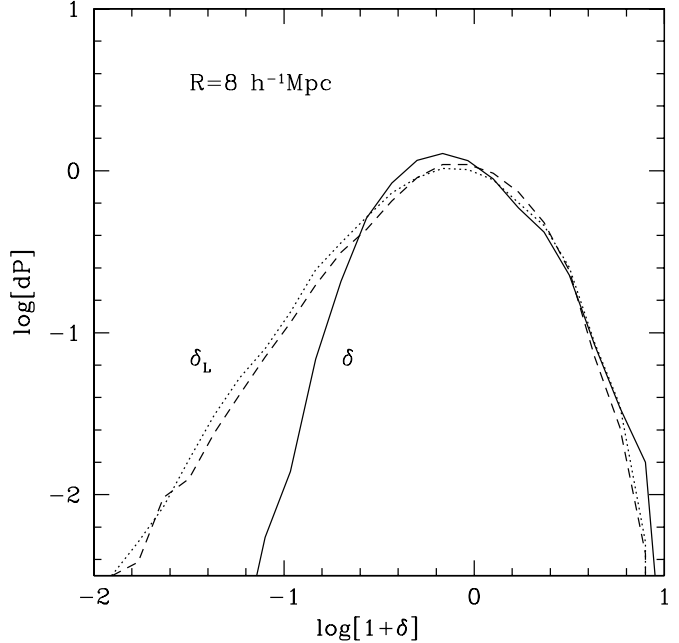,width=0.5\hdsize}}
\caption{The solid  and dashed curves show  the distribution functions
  of  the mass  overdensity  $\delta$ and  of  the luminosity  density
  $\delta_L$, respectively.  The dotted curve shows  the conversion of
  the distribution  of mass density to the  distribution of luminosity
  density  using  the   mean  $\delta_L$-$\delta$  relation  shown  in
  Fig.\,\ref{fig:dlasdm}.}
\label{fig:PofrhoL}
\end{figure}

In order  to show  that $\delta_L$  can be used  to rank  $\delta$, we
first fit the mean $\rho_L$-$\rho$ relation by the following function
\begin{equation}
\label{fitrel}
y=A e^{-x_*}x^{\beta} e^{x_*/x}\,,
\end{equation}
where  $x=1+\delta$, $y=1+\delta_L$,  and $A$,  $x_*$ and  $\beta$ are
fitting  parameters.  For high-density  cells ($x\gg  x_*$), $y\propto
x^{\beta}$,  while for  $x\ll x_*$,  $y$ decreases  exponentially with
decreasing $x$.   For $R=8\mpch$, we  obtain $A=1.0203$, $x_*=-0.3053$
and  $\beta=0.8035$. We  then use  this mean  relation to  convert the
distribution function of the mass density into a distribution function
for the luminosity density  and compare it with the luminosity-density
distribution function derived directly from the simulation. The result
is  shown in  Fig.~\ref{fig:PofrhoL}. Clearly,  the luminosity-density
distribution  function is  recovered  quite accurately  with the  mean
relation~(\ref{fitrel}), suggesting that the luminosity density can be
used to represent the mass density.
\begin{figure}
\centerline
{\psfig{figure=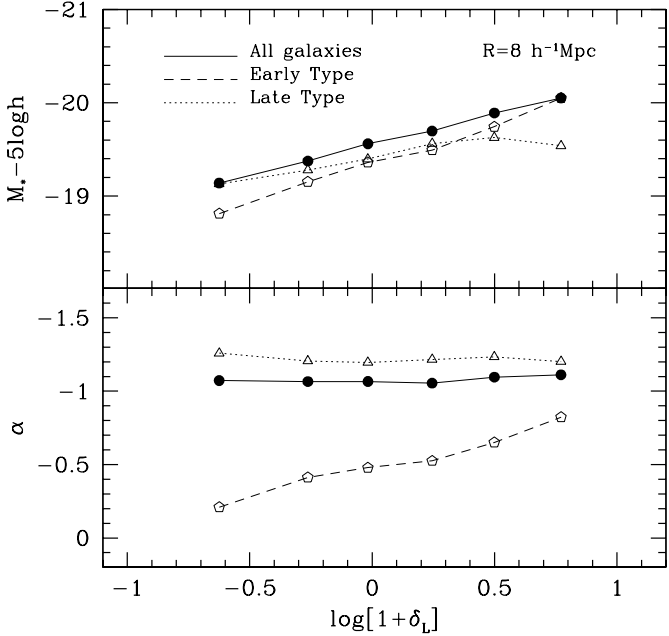,width=0.5\hdsize}}
\caption{The characteristic magnitude, $M_*$, and the faint-end slope,
  $\alpha$,  of  the  best-fit  Schechter  function  as  functions  of
  $\delta_L$.   The results are  shown for  the total,  early-type and
  late-type samples, as indicated.  Schechter functions are fit to the
  luminosity functions over the range $L \geq 10^{9}h^{-2} \Lsun$.}
\label{fig:LF_paraL}
\end{figure}
\begin{figure}
\centerline
{\psfig{figure=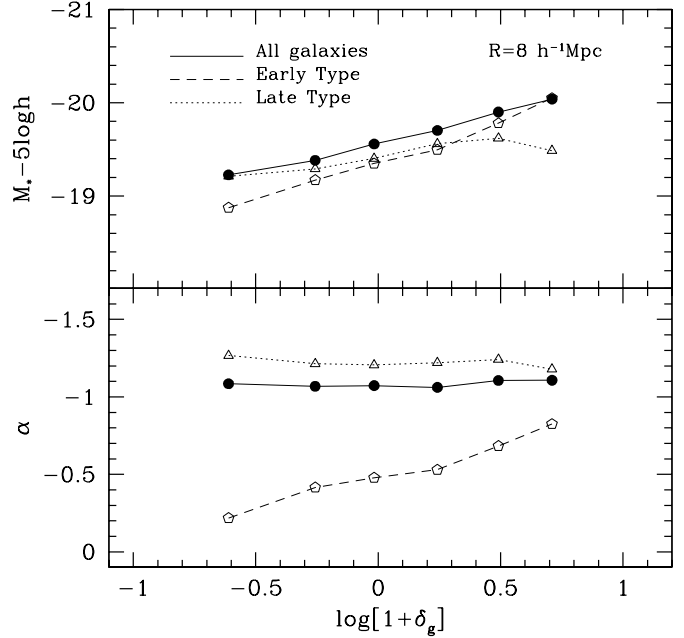,width=0.5\hdsize}}
\caption{Same  as  Fig.~\ref{fig:LF_paraL},   but  as  a  function  of
  $\delta_g$,  the overdensity (in  number) of  galaxies with  $L \geq
  10^{9}h^{-2} \Lsun$. }
\label{fig:LF_paraG}
\end{figure}

Fig.~\ref{fig:LF_paraL} shows how $L^{*}$ and $\alpha$ change with the
luminosity  density.   The  results  are  similar to  those  shown  in
Fig.~\ref{fig:LF_para},  except  that   the  range  of  $\delta_L$  is
stretched at the low-density end  relative to $\delta$, because of the
non-linear   relationship  between   $\delta_L$   and  $\delta$.    As
comparison,  we  also  show  $L^{*}$  and  $\alpha$  as  functions  of
$\delta_g$, where $\delta_g$  is based on the number  of galaxies with
$L\ge  10^9  h^{-2} \Lsun$  (Fig.\,\ref{fig:LF_paraG}).   With such  a
luminosity limit,  the results based  on $\delta_g$ are very  close to
those based on $\delta_L$.

\section{Discussion} 
\label{sec:discussion}

We have shown  that the halo occupation model,  which assumes that the
luminosity distribution in a dark halo only depends on its mass, makes
specific predictions about how  the galaxy LF changes with large-scale
environment.  The model predicts  that the  LFs for  relatively bright
galaxies  can  be  fit   reasonably  well  by  a  Schechter  function,
independent of environment. In all cases the characteristic luminosity
$L^{*}$ increases with local mass density. For late-type galaxies, the
faint-end slope is quite  independent of large-scale environment while
for early-types  the value  of $\alpha$ changes  from $\sim 0$  in low
density regions to $\sim -1$  in high-density regions. The predicted LF
for low-density regions shows significant departure from the Schechter
form  at   $L\la  10^9  h^{-2}\Lsun$,  and  this   departure  is  more
significant for early-type galaxies. All these predictions have simple
explanations  based   on  the  change  of  halo   mass  function  with
large-scale environment,  and the change of halo  occupation with halo
mass.

We have also  shown that the luminosity density  is tightly correlated
with the  mass density on large  scales. Thus, our  predictions can be
checked using large galaxy redshift  surveys. At the moment, the study
of the dependence of galaxy LF on local environment has mainly focused
on comparing  the LF of cluster  galaxies with that  of field galaxies
(see  e.g.   De Propris  \etal  2003  and  references therein).   More
related to the model predictions  presented here are the analyses made
by  H\"utsi \etal  (2002)  and Bromley  \etal  (1998).  H\"utsi  \etal
estimated galaxy LFs in spherical  volumes with a radius $10\mpch$ (in
redshift  space)   within  which  the  overdensities 
of galaxy number are respectively $\delta_g<0$, 
$0<\delta_g<1$,  and $\delta_g>1$, and found
that the  faint-end slope $\alpha$ is  about $-1.1$ for  all the three
cases, but  that the characteristic luminosity brightens  by about 0.4
magnitude  from  the  lowest-density  sample  to  the  highest-density
sample.   These results  are  consistent with  our  model prediction.  
Bromley \etal found  that the faint-end slope of  the LF of early-type
galaxies  (defined according  to  spectral type)  depends strongly  on
local  density, with  $\alpha$  increasing from  $\sim  -0.4$ in  high
density regions to  $\sim 0.2$ in low density  regions.  This trend is
also  consistent  with our  model  prediction. Unfortunately,  current
observational  results are  not yet  sufficiently accurate  to  give a
stringent constraint  on the model.   The situation will soon  change. 
With the use of large redshift surveys of galaxies, such as the 2dFGRS
and SDSS \footnote{During  the final stages of this  project, Hoyle et
  al (2003)  published a  paper based on  SDSS data that  adresses the
  environment-dependence of  the LF.  Their  results are qualitatively
  in  good agreement  with our  predictions,  though we  delay a  more
  detailed comparison  to a future  paper.}, one can estimate  to high
accuracy  the  galaxy luminosity  functions  in  regions of  different
luminosity  densities  or  galaxy  number densities
(Croton et al. in preperation), facilitating  an
accurate comparison between model and observations

 If our model prediction agrees in detail with observational 
data, it will add strong support to the assumptions made in the 
theory: the `concordance' cosmology (which determines the 
properties of the halo population) and that galaxy properties 
only depend on halo mass. If we assume that the halo population
in the real universe is not very different from that predicted
by the `concordance' cosmology, any significant difference 
between our model prediction and observation would have 
important implications for the formation of galaxies in the 
cosmic density field. For example, the prediction that the 
faint-end slope of the LF should not depend significantly 
on large-scale environment is due to the fact that most of the 
faint galaxies in our CLF model are in low-mass haloes. 
A significant discrepancy between model prediction and 
observation would therefore require a CLF model in which 
many faint galaxies are in massive haloes. This change of 
CLF is, however, strongly constrained by the observed 
weak clustering of faint galaxies. Thus, any significant 
discrepancy would strongly indicate the assumption
that the galaxy properties only depend on halo mass is incorrect, 
invalidating the whole halo occupation approach.
The other prediction of the model, that the characteristic 
luminosity increases mildly with local density, is due
to the fact that in our CLF model bright galaxies are 
preferentially found in massive halos. Here any discrepancy with 
observational data would therefore require a 
change in the fraction of bright galaxies in massive 
haloes. Such change will lead to a change in the correlation 
length as a function of luminosity, and is subject to sringent 
observational constraints. As mentioned above, the other CLF models 
we have obtained give very similar results, and so any significant 
discrepancy would mean that either galaxy properties depend 
significantly on other properties of halos in addition to the mass, 
or our parameterization of the CLF is not sufficiently general. 
As discussed in Paper I, our parameterization is 
motivated by physical considerations, and is quite general
as long as the CLF is assumed to be a smooth function of halo mass. 
Thus,  should a sigificant change in the CLF be required by
the observation, it would have a significant impact on our 
understanding of galaxy formation in dark matter haloes. 
 
As far as an accurate  comparison with observation is concerned, there
are several effects  to be taken into account.   The first is redshift
distortion. Since bright galaxies  in massive clusters are expected to
have large velocity  dispersion, some of these galaxies  may appear in
`low-density' regions  in redshift space, weakening  the dependence on
$\delta$. This  effect was seen  in our experiment where  $\delta$ was
defined in redshift space. The second is concerned with the definition
of  overdensity.   In a  magnitude-limited  sample,  volumes at  large
distances from us contain only bright galaxies, and so the estimate of
the overdensity in such a cell has to rely on a small number of bright
galaxies.   Since the shape  of the  luminosity function  changes with
$\delta$, the  correction of the  incompleteness due to  the magnitude
limit  in individual  volumes  must  take this  effect  into account.  
Finally,  since the  Schechter function  is not  a perfect  model, the
fitting parameters depend  on the luminosity {\it range}  used for the
fit. All these effects can be taken into account properly with the use
of mock catalogs  by applying the same analyses  to both the simulated
and observed samples.   We plan to present such  a detailed comparison
with observational data in a forthcoming paper.

It  is also  interesting to  compare the  predictions  presented here,
based  on the  CLF, with  those  based on  semi-analytical models  for
galaxy formation  (e.g. Kauffmann et  al. 1999; Somerville  \& Primack
1999; Benson et al. 2002; Mathis et al. 2002).  Mathis \& White (2002)
have shown that the characteristic luminosity of the galaxy luminosity
function in  their semi-analytical model  increases progressively from
low-density  to high-density  environments while  the  faint-end slope
becomes  slightly   shallower.   These  results   are  in  qualitative
agreement with  our predictions presented here.  As already emphasized
in  paper~II, this  suggests that  the halo  occupation  statistics as
described  by our  CLF nicely  fit  within the  standard framework  of
galaxy formation.

%%%%%%%%%%%%%%%
% Acknowledgments
%%%%%%%%%%%%%%%

\section*{Acknowledgement}
We  are  grateful to  Darren  Croton,  Glennys  Farrar, and  Guinevere
Kauffmann for useful discussions.   Numerical simulations used in this
paper were carried out at the Astronomical Data Analysis Center (ADAC)
of the National Astronomical Observatory, Japan.
YPJ is supported in part by NKBRSF (G19990754) and by NSFC.

%%%%%%%%%%%%%%%
% Bibliography
%%%%%%%%%%%%%%%

\label{lastpage}

\end{document}